\documentclass[pra,twocolumn,showpacs,superscriptaddress]{revtex4}
\usepackage{amsmath}
\usepackage{graphicx,graphics,psfrag,amsmath,calc,amssymb,mathrsfs}

\begin{document}

\title{Tomography of correlation functions for ultracold atoms via
time-of-flight images}
\author{Wei Zhang}
\affiliation{Department of Physics, Renmin University of China, Beijing, 100872 PR China}
\affiliation{FOCUS Center and MCTP, Department of Physics, University of Michigan, Ann
Arbor, MI 48109 USA}
\author{L.-M. Duan}
\affiliation{Department of Physics, University of Michigan, Ann
Arbor, MI 48109 USA}
\date{\today}

\begin{abstract}
We propose to utilize density distributions from a series of
time-of-flight images of an expanding cloud to reconstruct
single-particle correlation functions of trapped ultra-cold atoms.
In particular, we show how this technique can be used to detect
off-diagonal correlations of atoms in a quasi-one-dimensional
trap, where both real- and momentum- space correlations are
extracted at a quantitative level. The feasibility of this method
is analyzed with specific examples, taking into account finite
temporal and spatial resolutions in experiments.
\end{abstract}

\maketitle

%
%

\section{Introduction}

\label{sec:introduction}

Ultracold atomic gas provides a controllable platform to study
strongly correlated many-body physics, which has attracted strong
interest recently \cite{review}. To reveal many-body properties of
the underlying systems, normally one needs to detect some kinds of
correlation functions. The detection method for atomic systems is
typically very different from detection of condensed matter
materials. For solid state systems, linear responses provide a
useful method to measure the system correlation functions. For
ultracold atomic gases, measuring linear response is possible but
not always convenient. A powerful and widely used detection method
for atomic gases is based on the time-of-flight (TOF) imaging
technique~\cite{review}, which is unique to these atomic systems and
has no counterpart in condensed matter materials. During the TOF
imaging, one measures the light absorption of an expanding atomic
cloud released from the external trap. The light absorption gives
information of the density distribution of the expanded cloud,
which, under some approximation, is proportional to the initial
momentum distribution of the atomic gas prior to expansion.
Therefore, the TOF imaging provides us a useful technique to extract
the diagonal single-particle correlation function in the momentum
space. With only the diagonal correlation in the momentum space, in
general it is inadequate to reconstruct the real space correlation
function. To fulfill this gap, several methods have been proposed to
introduce additional control techniques, and measure the real space
correlation functions based on the atomic interference \cite
{dettmer-01} or the Fourier sampling \cite{duan}. Some of the
real-space correlations can also be inferred from measurement of the
two-particle correlations which can be extracted from noise
spectroscopy \cite{noise}.

In this manuscript, we discuss a method to extract information
about full single-particle correlations in both the momentum and
the real spaces, by measuring the density profiles in the TOF
images at different expansion times. Our detection proposal is in
the spirit of tomography in the spatial-temporal space, where the
full correlations are reconstructed through certain algorithms.
The method is strict for (quasi) one-dimensional (1D) systems, and
can be applied to higher-dimensional cases where the symmetry is
applied to reduce the effective dimensionality. Compared with
other detection schemes, this method is more direct in the sense
that it relies solely on the TOF images and does not require
introduction of other challenging control techniques. Besides, it
only utilizes the mean density distribution of the resulting
images, for which the detection is in general significantly easier
than the measurement of quantum noise of the corresponding images.
As an example of applications of this detection method, we
consider a quasi-1D Bose gas and demonstrate that all off-diagonal
single-particle correlations can be reconstructed from a series of
TOF images at various expansion times. The feasibility of this
method is analyzed under realistic experimental conditions of
spatial and temporal resolutions.

The remainder of this manuscript is organized as follows. In
Section \ref {sec:expansion}, we discuss the main reconstruction
formalism by analyzing the ballistic expansion process of an
atomic gas and showing how the single-particle correlations are
obtained from density profiles in the TOF images. We consider both
the quasi-1D case and the high-dimensional cases where one can
either separate variables or have spherical symmetry to reduce
dimensionality. As a by-product, we also derive in this section
the formula for the far-field limit, where the initial momentum
distribution is directly connected with the final density profile,
and obtain the quantitative conditions under which this
far-field-limit formula is valid. After introducing the general
formalism, to illustrate its applications we give an example in
Section~\ref{sec:1dBose} by considering a quasi-1D Bose gas, and
demonstrate how correlations are reconstructed from the TOF
images. Finally, we discuss the effects from finite spatial and
temporal resolutions and conclude that this method is applicable
within present experimental conditions.

\section{Formalism for reconstruction of correlation functions}

\label{sec:expansion}

In most experiments, the atomic gas is usually prepared inside
some external optical or magnetic trap. When the system achieves
its thermal equilibrium, one detects the properties of its
underlying many-body state through the TOF imaging. To perform
this TOF imaging measurement, one turns off the external trap such
that the atomic gas starts to expand in space. For simplicity, we
assume a ballistic expansion process where the atomic interaction
is negligible during the expansion. This is typically a good
approximation for optical lattice experiments when the atomic
density is not high \cite{note1}. For strongly interacting atoms
near a Feshbach resonance, to get the ballistic expansion, one
needs to first sweep the magnetic field to the deep BEC or BCS
side to turn off the interaction, as done in many experiments.

For a ballistic expansion, the atomic cloud freely expands and the
expansion dynamics is already known. Before turn-off of the trap,
the system is in thermal equilibrium, and its single-particle
correlation function is denoted by
$\mathcal{G}_{0}(\mathbf{r},\mathbf{r}^{\prime })=\langle \phi
_{0}^{\dagger }(\mathbf{r})\phi _{0}(\mathbf{r}^{\prime })\rangle
$, where $ \phi _{0}(\mathbf{r})\equiv \phi (\mathbf{r,}t\leq 0)$
is the atomic field operator, which is either bosonic or fermionic
corresponding to bosons or fermions, respectively. Assume at time
$t=0$ the trap is turned off, and we measure the density profile
$\langle n(\mathbf{r},t)\rangle \equiv \langle \phi ^{\dagger
}(\mathbf{r,}t)\phi (\mathbf{r},t)\rangle $ of the atomic cloud at
various expansion time $t$. From the measured density $\langle
n(\mathbf{r},t)\rangle $ \cite{note2}, we would like to
reconstruct the full single-particle correlation
$\mathcal{G}_{0}(\mathbf{r},\mathbf{r}^{\prime })$. Note that this
reconstruction is in general impossible for a three-dimensional
system, where $\mathcal{G}_{0}(\mathbf{r},\mathbf{r}^{\prime })$
depends on six variables and $\langle n(\mathbf{r},t)\rangle $ has
only four variables. The correlation
$\mathcal{G}_{0}(\mathbf{r},\mathbf{r}^{\prime })$ therefore
contains more information than the density profile $\langle
n(\mathbf{r},t)\rangle $. However, we will show in the following
that for any quasi-1D systems, the reconstruction of the
correlation functions can be done exactly. Furthermore, the
correlation can also be obtained for many practical
high-dimensional systems, where the system either possesses
spherical symmetry or can be separated in variables for the
single-particle correlations.

The expansion dynamics of the atomic cloud is well described by the
Schrodinger equation for the atomic field operator (taking $\hslash =1$):
\begin{equation}
i\partial _{t}\phi (\mathbf{r},t)=-\frac{\nabla ^{2}}{2m}\phi (\mathbf{r},t).
\label{eqn:schrodinger}
\end{equation}
The solution of this equation in the momentum space is simply given by
\begin{equation}
\phi (\mathbf{k},t)=\mathcal{U}_{0}(t)\phi _{0}(\mathbf{k})=\phi
_{0}( \mathbf{k}) e^{-i k^2 t/2m},
\label{eqn:freepropagator}
\end{equation}
where $\phi (\mathbf{k},t)$ is the Fourier transform of the field
operator $ \phi (\mathbf{r},t)$ at time $t$, $\mathcal{U}_{0}$ is
the free propagator, $m$ is the atomic mass, and initial conditions
are $\phi (\mathbf{k},t=0)=\phi _{0}(\mathbf{k})$. The expectation
value of density distribution thus takes the form
\begin{eqnarray}
\langle n(\mathbf{r},t)\rangle  &=&\frac{1}{\left( 2\pi \right)
^{6}}\iint d\mathbf{k}_{1}d\mathbf{k}_{2}e^{-i(\mathbf{k}_{1}-\mathbf{k}_{2})\cdot
\mathbf{r}}  \notag  \label{eqn:density} \\
&&\times e^{i(\mathbf{k}_{1}^{2}-\mathbf{k}_{2}^{2})t/2m}
\langle \phi _{0}^{\dagger }(\mathbf{k}_{1})\phi _{0}(\mathbf{k}%
_{2})\rangle .
\end{eqnarray}%
It is useful to define new variables $\mathbf{k}_{+} \equiv
\mathbf{k}_{1} + \mathbf{k}_{2}$ and $\mathbf{k}_{-} \equiv
\mathbf{k}_{1} - \mathbf{k}_{2}$, with which the density
expectation value becomes
\begin{eqnarray}
\langle n(\mathbf{r},t)\rangle  &=&\frac{1}{2\left( 2\pi \right)
^{6}}\iint d \mathbf{k}_{+} d\mathbf{k}_{-} e^{
-i\mathbf{k}_{-}\cdot \left( \mathbf{r}/\tilde{t} -\mathbf{k}_{+}
\right) {\tilde{t}}} \notag \label{eqn:density2}
\\
&&\times \left\langle \phi _{0}^{\dagger }\left( \frac{\mathbf{k}_{+}+%
\mathbf{k}_{-}}{2}\right) \phi _{0}\left( \frac{\mathbf{k}_{+}-\mathbf{k}_{-}%
}{2}\right) \right\rangle ,
\end{eqnarray}%
where ${\tilde{t}}\equiv t/2m$ is defined to simplify the notation.

An important feature of the density profile in Eq.
(\ref{eqn:density2}) is that for long evolution time $\tilde t$
(called the far-field-limit), the exponential term gives a
rapid-oscillating phase factor except for the
region around the point $(\mathbf{k}_{-}=0,\mathbf{k}_{+}=\mathbf{r}/{\tilde{%
t}})$. As a consequence, the integration over $\mathbf{k}_{+}$ and $\mathbf{k%
}_{-}$ is dominated by the contribution from such region, leading
to an approximating form for the density expectation value (see
the derivation in Appendix A)
\begin{equation}
\langle n(\mathbf{r},t)\rangle \approx \frac{1}{2\left( 2\pi \right) ^{3}{%
\tilde{t}}}\left\langle \phi _{0}^{\dagger }\left( \frac{\mathbf{r}}{2\tilde{%
t}}\right) \phi _{0}\left( \frac{\mathbf{r}}{2\tilde{t}}\right)
\right\rangle .  \label{eqn:density-FFL}
\end{equation}%
Notice that the final density profile is directly proportional to
the initial momentum distribution via a scaling relation
$\mathbf{k}=\mathbf{r} /(2\tilde{t})$. This formula has been
widely used for interpretation to the measurement result from the
TOF images. For this far-field-limit to be valid, the expansion
time needs to be sufficiently long. Quantitatively, it has to
satisfy the following condition
\begin{equation}
\left( \Delta \mathbf{k}_{+}\cdot \Delta \mathbf{k}_{-}\right) \tilde{t}\gg
1.  \label{eqn:FFL}
\end{equation}%
Here, $\Delta \mathbf{k}_{+}$ and $\Delta \mathbf{k}_{-}$ are
characteristic scales for the extension (variation) of
single-particle correlation function
$\mathcal{G}_{0}(\mathbf{k}_{1},\mathbf{k}_{2})$ along the
$\mathbf{k}_{+}$ and $\mathbf{k}_{-}$ directions, respectively. The
derivation of this condition for the far-field limit is shown in
Appendix \ref{app:farfield}.

The extraction of momentum distribution only utilizes the measured
density profile in the far-field limit. The density distribution
measured at other expansion times contain more information about the
single-particle correlations. Next, we discuss how to inverse the
relation in Eq.~(\ref{eqn:density}), and to reconstruct the
off-diagonal correlations from TOF images. We first discuss in
detail the 1D system where the reconstruction algorithm is exact,
and then generalize the method to higher dimensional cases by
reducing to a set of 1D problems when symmetry or separability
arguments are applicable.

Let us consider a quasi-1D atomic cloud along the $x$-direction
where the transverse degrees of freedom are frozen through a deep
potential such as an optical lattice. The density profile
$\left\langle n(x,t)\right\rangle $ thus depends on correlation
functions through a double-integration with two phase factors
along the spatial and temporal directions, respectively. One can
think about performing a double Fourier transform to inverse the
relation. Specifically, we obtain%
\begin{eqnarray}
\langle {\tilde{n}}(p,\omega )\rangle  &\equiv &\int_{-\infty }^{\infty
}dx\int_{0}^{\infty }dt\langle n(x,t)\rangle e^{-i\left( px+\omega t\right) }
\notag \\
&&\hspace{-2cm}=\frac{m}{2|p|}\left\langle \phi _{0}^{\dagger }\left( \frac{%
-2m\omega /p-p}{2}\right) \phi _{0}\left( \frac{-2m\omega /p+p}{2}\right)
\right\rangle   \notag \\
&&\hspace{-2cm}-\frac{i}{4\pi }\frac{2m}{p}\mathcal{PV}\left( \int dk_{+}%
\frac{\left\langle \phi _{0}^{\dagger }\left( \frac{k_{+}-p}{2}\right) \phi
_{0}\left( \frac{k_{+}+p}{2}\right) \right\rangle }{-2m\omega /p+k_{+}}%
\right) \label{eqn:n(p,w)}
\end{eqnarray}%
for the cases with $p\neq 0$, which is connected with the off-diagonal
momentum correlations and hence of particular interest.

Notice that the integration over time is for $t>0$ only, since all
the density profiles are obtained \textit{after} turn-off of the
trap. This partial Fourier transform hence generates the principal
value ($\mathcal{PV}$) integral on the right hand side of
Eq.~(\ref{eqn:n(p,w)}). This expression can be significantly
simplified if the single-particle correlations are real. This
condition is equivalent to assume a time-reversal symmetry
throughout the system. In fact, for systems in a stationary frame,
it is very unlikely for the time-reversal symmetry to be broken as
long as there is no such symmetry breaking term in the Hamiltonian
\cite{note3}.  Under this condition, the principal value integral
is purely imaginary, and the off-diagonal momentum correlation
(with $k_{-}\neq 0$) can be solved as
\begin{eqnarray}
&&\left\langle \phi _{0}^{\dagger }(k_{1})\phi _{0}(k_{2})
\right\rangle \nonumber \\ \label{eqn:Kcorrelation}
&=&4|k_{-}|\mathrm{Re}\left[ \int_{0}^{\infty} d{\tilde{t}}
\int_{-\infty}^{\infty} dx e^{-i k_- (k_+ {\tilde{t}} -x)}
\left\langle n(x,t)\right\rangle \right] .
\end{eqnarray}%
Therefore, with the measured density profile $\left\langle
n(x,t)\right\rangle $ at different time $t$, we can reconstruct all
the momentum correlations through the formulae
(\ref{eqn:density-FFL}) and (\ref{eqn:Kcorrelation}). With the
knowledge of all momentum space correlations, the real space
correlation function can also be obtained via a double Fourier
transform. After substituting the expression
(\ref{eqn:Kcorrelation}), the integration over two momentum indices
can be carried out analytically,
leading to%
\begin{eqnarray}
&&\left\langle \phi _{0}^{\dagger }(x_{1})\phi _{0}(x_{2})\right\rangle
\notag  \label{eqn:Rcorrelation} \\
&=&\frac{1}{2\pi }\mathrm{Re}\left[ \int_{0}^{\infty }d{\tilde{t}}%
\int_{-\infty }^{\infty }dx
\left\vert \frac{x_{-}}{{\tilde{t}}^{2}}\right\vert
e^{-ix_{-}\left( 2 x - x_{+}\right) / 4{\tilde{t}} }
\left\langle n(x,t) \right\rangle \right] .
\end{eqnarray}%
Here, the variables $x_{+}\equiv x_{1}+x_{2}$ and $x_{-}\equiv
x_{1}-x_{2}$ are defined to simplify notations. The convergence of
the integration over time for $t\rightarrow 0$ is usually guaranteed
by cancelation from the fast oscillation in the exponent
$e^{-ix_{-}\left( 2x - x_{+}\right) /4{\tilde{t}} }$. These two
simple relations between the measured density profiles and the
initial correlation functions in both the momentum space [Eq.
(\ref{eqn:Kcorrelation})] and the real space [Eq.
(\ref{eqn:Rcorrelation})] are the central results for this section.

After introducing the reconstruction scheme for the 1D case, next we
consider the problem in higher dimensions. Since the density profile $%
\left\langle n(\mathbf{r},t)\right\rangle $ is in fact a function of $d$%
(spatial)$+1$(temporal) variables, in the most general case it is no
longer possible to fully reproduce the single-particle correlation,
which is a function of $d\times d$ variables. However, if the system
has some properties which allows us to reduce dimensionality, a
similar procedure can still be performed on the reduced problems.
For instance, if there exists spherical symmetry in 3D systems, the
field operator can be written as $\phi (\mathbf{r},t)=\phi (r,t)$,
where $r$ is the radial distance. In this case, the free expansion
process is governed by the Schrodinger equation in spherical coordinate,
\begin{equation}
i\partial _{t}\phi (r,t)=-\frac{1}{2m}\left( \partial _{r}^{2}+\frac{2}{r}%
\partial _{r}\right) \phi (r,t).  \label{eqn:schrodinger-3D}
\end{equation}%
By defining a new operator $\psi (r,t)=r\phi (r,t)$, the equation above
becomes %
\begin{equation}
i\partial _{t}\psi (r,t)=-\frac{\partial _{r}^{2}}{2m}\psi (r,t).
\label{eqn:schrodinger-3D-2}
\end{equation}%
Notice that this equation has exactly the same structure as the Schrodinger
equation (\ref{eqn:schrodinger}) for the 1D case, which allows us to obtain
the single particle correlation function $\langle \phi _{0}^{\dagger
}(r)\phi _{0}(r^{\prime })\rangle =\langle \psi _{0}^{\dagger }(r)\psi
_{0}(r^{\prime })\rangle /(rr^{\prime })$ from the density distribution $%
n(r,t)$ via a similar procedure as Eq. (\ref{eqn:Rcorrelation}).

The reconstruction scheme can also be generalized to higher dimensional
cases when the correlation functions are separable, i.e.,
\begin{equation}
\left \langle \phi^{\dagger }(\mathbf{r},t)
\phi(\mathbf{r}^{\prime},t) \right \rangle = \prod_{i=x,y,z} \left
\langle \phi_{i}^{\dagger }(r_{i},t)\phi _{i}(r_{i}^{\prime },t)
\right \rangle. \label{eqn:separable}
\end{equation}%
In this case, the higher dimensional problem can be effectively reduced to a
set of 1D problems and then treated separately using the same method
discussed above.

\section{Correlations in one-dimensional quasi-condensates: an example}

\label{sec:1dBose}

In the previous section, we have shown how to reconstruct the
single-particle correlations from the measured density profile for
1D systems and higher-dimensional cases where symmetry or
separability can be used to reduce the effective dimensionality.
To illustrate applications of this general formalism, in this
section we consider a specific example and show hot to reconstruct
the correlation functions for bosonic atoms in a quasi-1D trap.
The reconstruction formula in Eqs. (\ref{eqn:Kcorrelation}) and
(\ref {eqn:Rcorrelation}) are given as double integrals of the
measured density profiles. In realistic experiments, however, one
can take only a finite number of images, and each image contains a
finite number of resolvable data points. In the following
discussion, we will take into account the finite spatial and
temporal resolutions, and show that a good approximation to the
correlation function can be inferred already from about a dozen of
images.

In this example, we consider a quasi-1D Bose gas trapped through
either a highly elongated cigar-shaped potential \cite{gorlitz-01}
or a deep transverse two-dimensional optical
lattice~\cite{kinoshita-04}. In this quasi-1D geometry with strong
radial confinement $\omega _{\rho}$, atoms at low temperatures are
essentially frozen at the ground state of the radial harmonic trap.
Thus the radial wave function is governed by
\begin{equation}
\phi _{\rho }(y,z)=\sqrt{\frac{1}{\pi a_{\rho
}^{2}}}e^{-(y^{2}+z^{2})/(2a_{ \rho }^{2})},  \label{eqn:radial}
\end{equation}%
where $a_{\rho }\equiv 1/\sqrt{m\omega _{\rho }}$ characterizes the
extension of the radial wave function and $m$ is the atomic mass.
When $a_{\rho }$ is much greater than the effective length scale of
interatomic potential $R_{e}$, the gas can be described with an
effective 1D interaction rate~\cite{1Dgas}
\begin{equation}
g_{\mathrm{1D}}=\frac{2a_{s}}{ma_{\rho }^{2}},  \label{eqn:1dinteraction}
\end{equation}%
where $a_{s}$ denotes the $s$-wave atomic scattering length in free
space.

The static single-particle correlation function in this quasi-1D system has
been calculated by evaluating phase fluctuation effects around the saddle
point condensate densities. This process gives the axial real-space
correlation function~\cite{1Dgas}%
\begin{equation}
\mathcal{G}_{0}(x,x^{\prime })
\approx \sqrt{n_{0}(x)n_{0}(x^{\prime })}%
e^{-1/2F_{s}(x,x\prime )},  \label{eqn:1dcorrelation}
\end{equation}%
where $n_{0}(x)$ is the zero-temperature Thomas-Fermi density distribution
in the axial direction%
\begin{equation}
n_{0}(x)=\frac{1}{g_{\mathrm{1D}}}\left( \mu_{0} -
\frac{m\omega_{x}^{2}x^{2}}{2} \right) \theta \left( \mu_{0} -
\frac{m\omega_{x}^{2}x^{2}}{2} \right) , \label{eqn:TFdensity}
\end{equation}%
and the function $F_{s}(x,x^{\prime })$ takes the form%
\begin{equation}
F_{s}(x,x^{\prime })=\frac{4T\mu _{0}}{3N\omega _{x}^{2}}\left\vert \ln %
\frac{(1-x/R_{\mathrm{TF}})(1+x^{\prime }/R_{\mathrm{TF}})}{(1+x/R_{%
\mathrm{TF}})(1-x^{\prime }/R_{\mathrm{TF}})} \right\vert .
\label{eqn:Ffunction}
\end{equation}%
Here, $T$ is the temperature, $N$ is the total atom number,
$\mu_{0}$ denotes the chemical potential at the trap center, and
$R_{\mathrm{TF}}=\sqrt{2\mu _{0}/(m\omega _{x}^{2})}$ is the
Thomas-Fermi cloud radius along the axial direction (with a small
axial trapping frequency $\omega _{x}$). We emphasize that we only
use the correlation function in Eq.~(\ref {eqn:1dcorrelation}) as an
example to illustrate our detection method. The derivation of this
correlation is irrelevant for our following purpose.

\begin{figure}[tbp]
\begin{center}
\includegraphics[width=8.0cm]{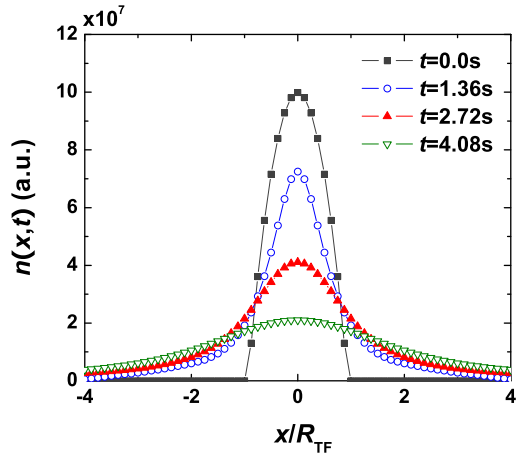}
\end{center}
\caption{ (Color online) The density profiles along the axial
direction at different expansion times for a quasi-1D Bose gas
released from the trap. Throughout the paper, we consider a gas of
$^{87}$Rb atoms with total number $N=10^{4}$ interacts via s-wave
scattering length $a_{s}=5.45$ nm. The gas is
confined in a cigar-shaped potential with the trapping frequencies $\protect\omega %
_{x}=2\protect\pi \times 30$ Hz and $\protect\omega _{\protect\rho }=2%
\protect\pi \times 20000$ Hz. The Thomas-Fermi radius along the
axial direction is $R_{\mathrm{TF}}\approx 160\protect\mu $m. Such
a system can be realized in a 2D optical lattice with the lattice
constant $d=426$ nm and the lattice depth
$V_{0}=10E_{\mathrm{R}}$, where $E_{\mathrm{R}}$ is the recoil
energy. In this plot, we consider a spatial resolution of $\Delta x=20%
\protect\mu $m, which gives about $16$ data points (squares) from
the absorption image at the very beginning of $t=0$. The
temperature of the system is assumed to be $T=20\protect\omega
_{x}$. } \label{fig:TOF}
\end{figure}

We assume a ballistic expansion of the atomic gas along the axial
direction for TOF imaging (the transverse trap is still on), which
allows us to numerically simulate the expansion process with the
initial correlation given by Eq.~(\ref{eqn:1dcorrelation}), and
obtain the expected density distribution at each measurement time.
Using these resulting density profiles, then we try to reconstruct
the corresponding correlation functions using Eqs.
(\ref{eqn:density-FFL}) and (\ref{eqn:Kcorrelation}). From the
derivation in the last section, if we have infinite spatial
resolution and take an infinite number of images, we should be able
to obtain exactly the same correlation function as shown in Eq.
(\ref{eqn:1dcorrelation}). So the purpose here is actually to
analyze the effects from finite spatial and temporal resolutions as
is the case for realistic experiments. In the following discussion,
we assume to take only $10 \sim 30$ images, and for each image we
only know $\left\langle n(x,t)\right\rangle $ for a discrete set of
points from the finite spatial resolution. In Fig.~\ref{fig:TOF}, we
show a typical set of results for $\left\langle n(x,t)\right\rangle
$ at various times. Notice that we assume here a finite spatial
resolution of $\Delta x=20\mu $m with $R_{\mathrm{TF}}/\Delta
x\approx 8$, which gives about $16$ readable data points from the
absorption image at the very beginning of $t=0$.

\begin{figure}[tbp]
\begin{center}
\includegraphics[width=8.0cm]{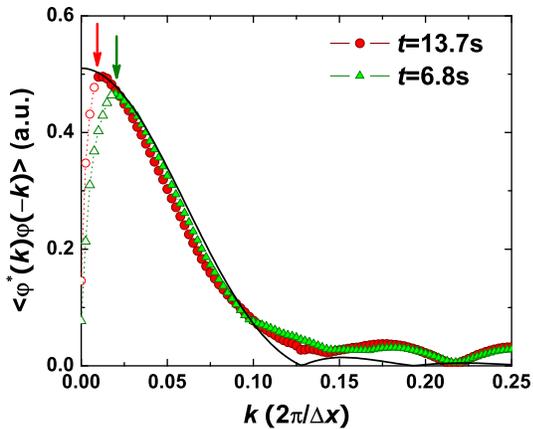}
\end{center}
\caption{ (Color online) The momentum-space correlation $\langle
\protect\phi_{0}^{\dagger }(k)\protect\phi _{0}(-k) \rangle$
reconstructed from the density profiles via
Eq.~(\protect\ref{eqn:Kcorrelation}), where a total of $M_{t}=30$
images are taken equidistantly between $t=0$ and the maximal
expansion time $t_{\mathrm{max}}=6.8$ s (triangles) and
$t_{\mathrm{max}}=13.7$ s (dots). The results are compared with the
expected exact values from a direct Fourier transform of
Eq.~(\ref{eqn:1dcorrelation}) (solid line). The oscillatory behavior
for large $k$ is an artificial effect due to the sudden drop of the
Thomas-Fermi density distribution at the cloud edge. Notice that the
reconstruction fails for $k\rightarrow 0$ (hollow points), while the
questionable region shrinks (arrows) with increasing
$t_{\mathrm{max}}$ and $x_{\mathrm{max}}$. In this plot, we assume a
spatial resolution of $\Delta x=40\protect\mu $m with
$R_{\mathrm{TF}}/\Delta x=4$, which gives about $8$ data points from
the absorption image at the very beginning of $t=0$. Other
parameters used here are the same as the ones in Fig.~\protect
\ref{fig:TOF}. } \label{fig:Kcorrelation}
\end{figure}

First, let us try to extract the correlation function in
momentum space using the formula in Eq.~(\ref{eqn:Kcorrelation}). In
Fig.~\ref{fig:Kcorrelation}, we plot the reconstructed off-diagonal
correlation $ \langle \phi _{0}^{\dagger }(k)\phi _{0}(-k) \rangle $
using the simulated density profiles. There are three important
features one can read from this plot. First, the reconstructed
correlations are fairly close to the expected exact values obtained
from the Fourier transform of the real-space correlation in
Eq.~(\ref{eqn:1dcorrelation}), especially when the correlation is
sizable such that the error is relatively small. Second, the
reconstruction fails for $k$ very close to zero. This is because the
TOF images taken here are not for an infinite time duration. For the
reconstruction, we take the data points within the ranges given by
$x_{ \mathrm{max}}$ and $t_{\mathrm{max}}$ for coordinates and time,
respectively. Since the reconstruction relies on a Fourier
transform, these finite ranges set a limit for the resolution in the
momentum space. In fact, by increasing the evolving time
$t_{\mathrm{max}}$, which simultaneously requires an increase of
$x_{\mathrm{max}}$ since the cloud expands more, we can push the
reconstruction further towards $k=0$. Third, except for the region
$k\sim 0$, the results are insensitive to the time split between two
subsequent images. Notice that since we use $M_{t}=30$ images for
both reconstructions in Fig. \ref{fig:Kcorrelation}, the trial with
longer evolving time has larger time split. The result, however, is
fairly close to
the other trial for $k$ not very close to $0$. %
\begin{figure}[tbp]
\begin{center}
\includegraphics[width=9.0cm]{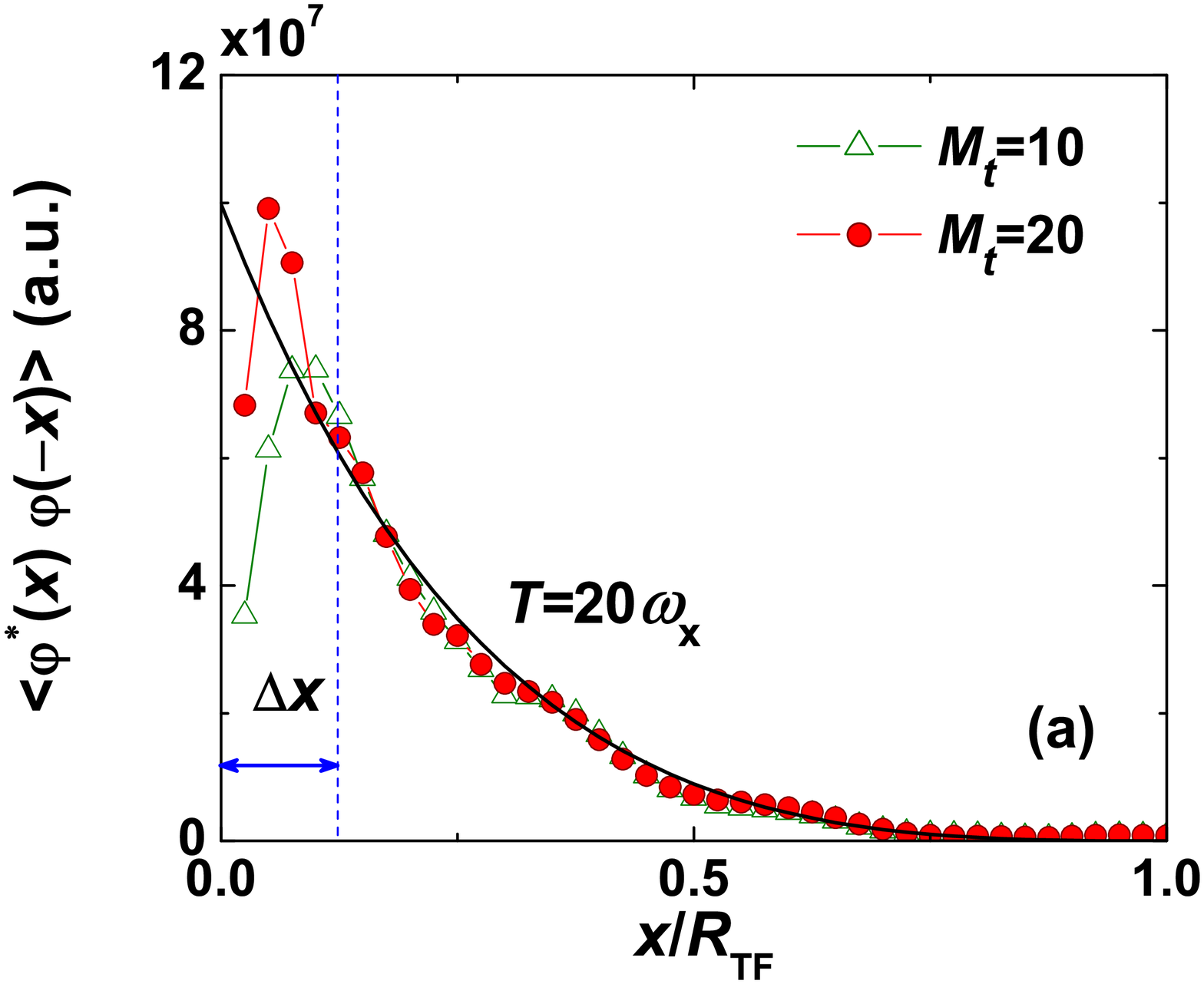} %
\includegraphics[width=9.0cm]{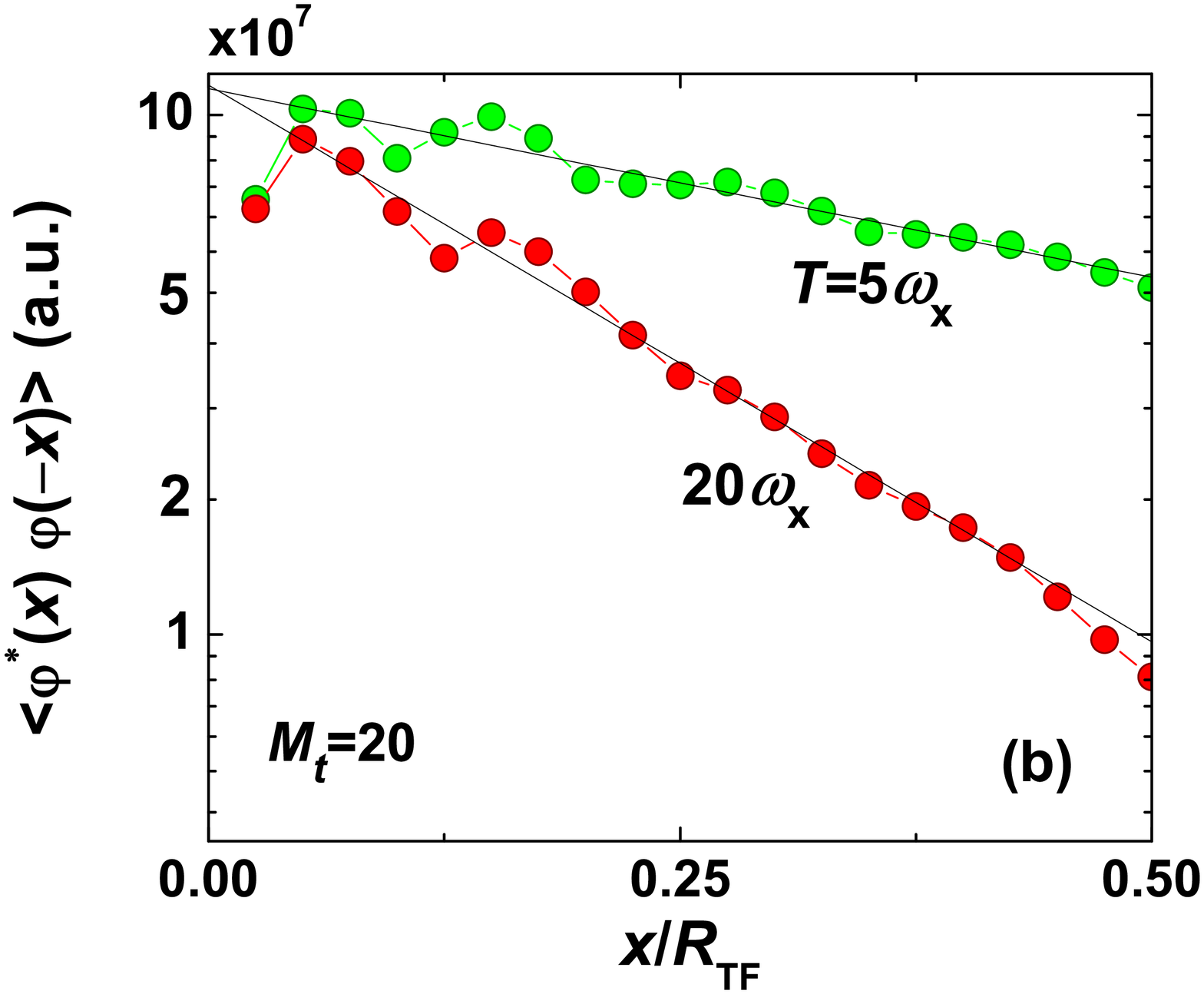}
\end{center}
\caption{(Color online) (a) The real-space correlation $\langle
\protect\phi_{0}^{\dagger }(x)\protect\phi _{0}(-x) \rangle$
reconstructed from the density profiles via
Eq.~(\protect\ref{eqn:Rcorrelation}), where the maximal expansion time $t_{%
\mathrm{max}}=6.8$ s are sliced equidistantly to obtain $M_{t}=10$
(triangles) and $M_{t}=20$ (dots) images. Results are compared with
the expected values of Eq.~(\protect\ref{eqn:1dcorrelation}) (solid
line). (b) Log plot of $\langle \phi_{0}^{\dagger}(x) \phi_{0}(-x)
\rangle$ around the center of the cloud at different temperatures.
Exponential decay can be clearly observed in this plot, with the
correlation length given by an exponential fit (solid lines). Other
parameters used here are the same as the ones in
Fig.~\protect\ref{fig:Kcorrelation}.} \label{fig:Rcorrelation}
\end{figure}

Next, we analyze the correlation function in the real space. In
Fig.~\ref{fig:Rcorrelation}(a), we show the reconstructed
correlations of $\langle \phi _{0}^{\dagger }(x)\phi_{0}(-x)\rangle
$ from the density profiles via Eq.~(\ref {eqn:Rcorrelation}). The
most striking feature of this plot is that the spatial correlations
can be obtained very precisely for $x$ not too close to $0$, using
as few as $M_{t}=10$ images. Even in the questionable region for $x$
less than the spatial resolution $\Delta x$, some information can
still be extracted by applying linear interpolation between image
pixels, and the precision in that region can be significantly
enhanced when more images are used for reconstruction. In
Fig.~\ref{fig:Rcorrelation}(b), the real-space correlations for two
different temperatures are plotted in log scale, showing explicitly
the exponential decay around the center of the trap with
corresponding correlation lengths.

\begin{figure}[tbp]
\begin{center}
\includegraphics[width=8.0cm]{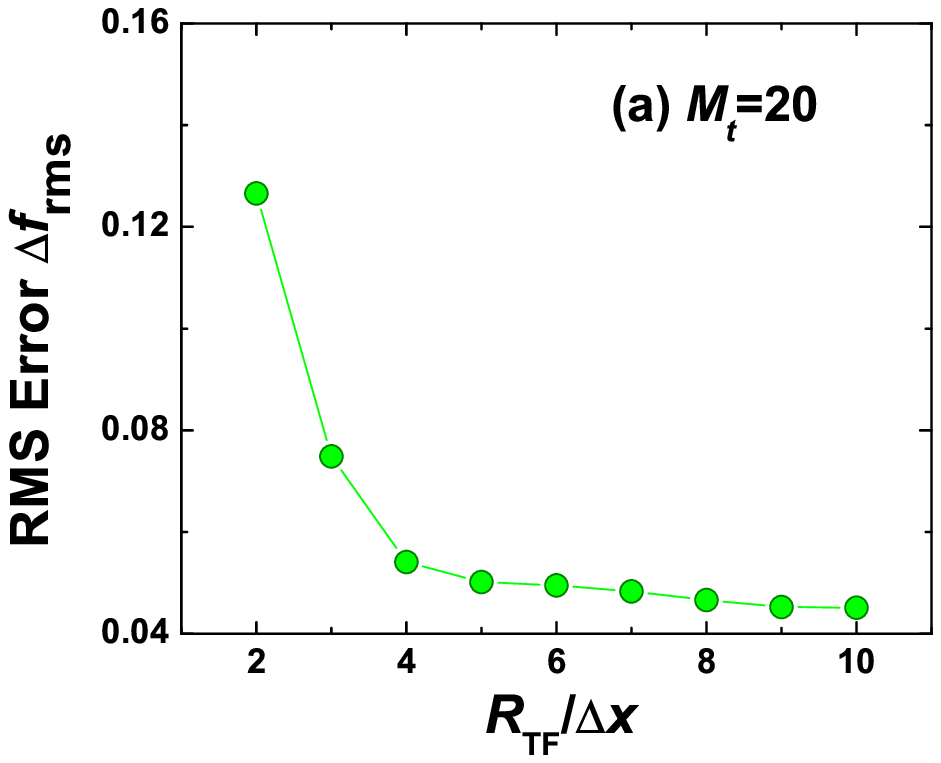} %
\includegraphics[width=8.0cm]{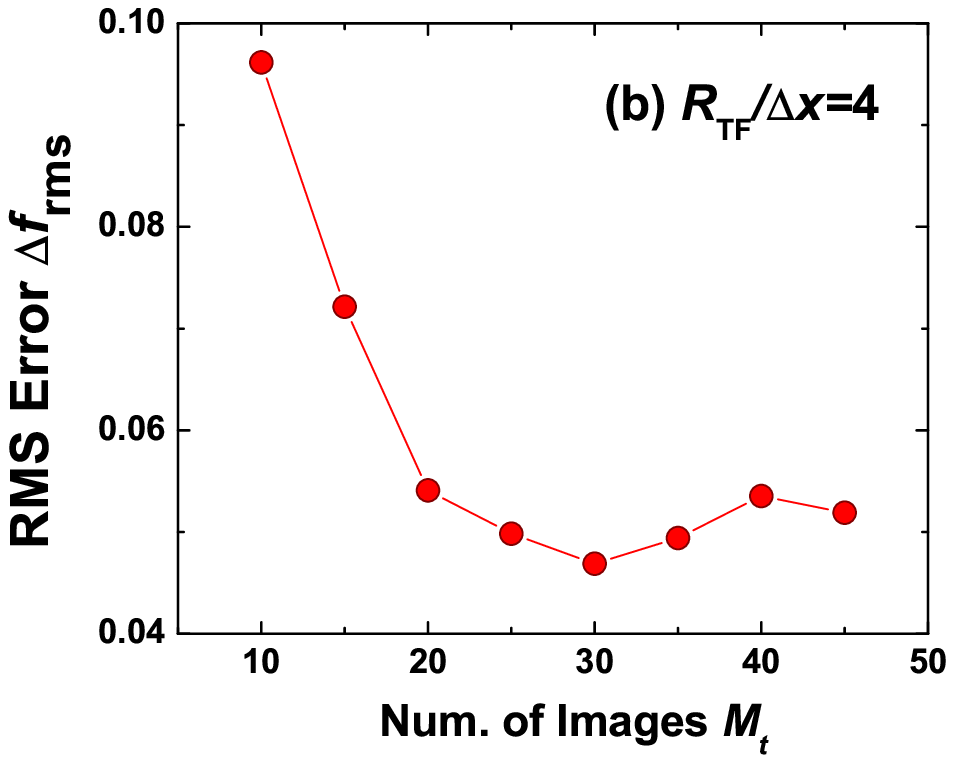}
\end{center}
\caption{ (Color online) Normalized RMS error of the reconstructed
correlations for $\mathcal{G}_{0}(x,-x)$, (a) as a function of the
spatial resolution $\Delta x$ with a fixed number of images
$M_{t}=20$, and (b) as a function of the number of images $M_{t}$
with a fixed spatial resolution $R_{\mathrm{TF}}/\Delta x=4 $. The
system is at temperature $T=20\protect\omega _{x}$, with total
expansion time $t_{\mathrm{max}}=6.8$ s. Other parameters are the
same as the ones in Fig.~\protect\ref{fig:TOF}. }
\label{fig:error}
\end{figure}

Finally, we investigate how the increase of spatial and temporal
resolutions can enhance the reconstruction precision. To
quantitatively evaluate the precision, we consider the real-space
correlation as illustrated in Fig.~\ref {fig:Rcorrelation}(a), and
define the root-mean-square (RMS) of the error
\begin{equation}
\Delta f_{\mathrm{rms}}=\frac{1}{n_{0}(0)}\sqrt{\frac{\sum_{i=1}^{N_{x}}[%
\mathcal{G}_{0}(x_{i},-x_{i})-\mathcal{G}_{0}^{\mathrm{TOF}%
}(x_{i},-x_{i})]^{2}}{N_{x}}},  \label{eqn:RMSerror}
\end{equation}%
where $\mathcal{G}_{0}^{\mathrm{TOF}}(x_{i},-x_{i})$ is the
reconstructed correlation from the TOF images, and the function is
normalized to the center density $n_{0}(0)$. Notice that this
normalized RMS error will depend on the number of reconstruction
sampling points $N_{x}$, which is $40 $ as in
Fig.~\ref{fig:Rcorrelation}(a). But its value will not change much
with different $N_{x}$ and will follow the same trend with variation
of the resolutions. In Fig.~\ref{fig:error}, we show the normalized
RMS error in Eq.~(\ref{eqn:RMSerror}) as functions of spatial
[Fig.~\ref{fig:error}(a)] and temporal [Fig.~\ref{fig:error}(b)]
resolutions, respectively. Notice that the error decreases rapidly
by increasing resolutions in both variables, as one would expect.
Besides, the value of the error saturates to a fairly small number
$(\sim 5\%)$ around $R_{\mathrm{TF}}/\Delta x\sim 4$ and $M_{t}\sim
20$, respectively, indicating that the technique is quite feasible
under the present technology.

\section{Conclusion}

\label{sec:conclusions}

In summary, we have proposed a method to reconstruct the full static
single-particle correlation functions in both the momentum and the real
spaces for cold atomic gas by measuring the density profiles at different
expansion times with the time-of-flight imaging. The method applies to
quasi-1D systems and can be generalized to higher dimensions when symmetry
or separability arguments can be used to reduce the effective
dimensionality. As an example, we consider a quasi-1D Bose gas and
demonstrate how real- and momentum-space correlations are reconstructed at a
quantitative level. The feasibility of this method is analyzed by evaluating
the reconstruction error with various spatial and temporal resolutions, and
the result suggests that the correlations can be inferred with pretty good
precision already with a dozen of images at practical spatial resolution.

\begin{acknowledgements}

This work was supported by the AFOSR\ through MURI, the DARPA, and
the IARPA.

\end{acknowledgements}

\appendix

\section{Far-field Limit}

\label{app:farfield}

For simplicity of the notation, here we derive the far-field limit
formula only for the 1D case. The extension of the formula to higher
dimensions is straightforward. In the 1D case, the expectation value
of the density distribution in Eq. (\ref{eqn:density2}) takes the
following form
\begin{equation}
I=\frac{1}{2\left( 2\pi \right) ^{2}}\int dx\int dye^{ixyt}f(x,y),
\label{eqn:app1-density}
\end{equation}
where $f(x,y)$ represents the single-particle correlation
function. The correlation function $f(x,y)$ spreads over certain
ranges, with the characteristic length scales along the $x$ and
$y$ directions denoted by $\Delta X$ and $\Delta Y$, respectively.
The integration over $dy$ in Eq. (\ref{eqn:app1-density}) can be
performed after a Fourier transform, leading to
\begin{eqnarray}
I &=&\frac{1}{4\pi }\int dx\mathscr{F}(x,s=xt)  \notag
\label{eqn:app1-density2} \\
&=&\frac{1}{4\pi t}\int ds\mathscr{F}(s/t,s),
\end{eqnarray}%
where $\mathscr{F}(x,s)$ is the Fourier transform of $f(x,y)$ as a
function of $y$. Notice that from the Fourier transformation
theorem, the function $\mathscr{F}$ must spread along the $x$ and
$s$ directions with the characteristic length scales given by
$\Delta X$ and $1/\Delta Y$, respectively.

In the expression above, the integration over $s$ is essentially along the
line $(s/t,s)$. For large enough time $t$, this line is almost the $s$-axis.
Thus, the result can be approximated by%
\begin{equation}
I\approx \frac{1}{4\pi t}\int ds\mathscr{F}(0,s)=\frac{1}{4\pi t}f(0,0),
\label{eqn:app1-density3}
\end{equation}%
where the second step is given by Fourier transforming the
$\mathscr{F}$ function back to the $(x,y)$ plane. The last equation
gives the far-field limit result.

In order to make this approximation valid, the function
$\mathscr{F}(s/t,s)$ and $\mathscr{F}(0,s)$ should be close to each
other for typical values of $s$. As the function
$\mathscr{F}(s/t,s)$ has characteristic length scales $\Delta X$ and
$1/\Delta Y$ respectively along the $x$ and $s$ directions, the
condition $s/t\ll \Delta X$ must be fulfilled for $s$ within the
typical range given by $1/\Delta Y$. Therefore, we conclude that the
far-field limit given by Eq. (\ref{eqn:app1-density3}) is valid when
the following condition is satisfied
\begin{equation}
\Delta X\Delta Yt\gg 1.  \label{eqn:app1-far-field}
\end{equation}%
The above argument can be easily generalized to higher dimensional
cases, leading to the far-field limit condition as shown in Eq.
(\ref{eqn:FFL}).


\begin{thebibliography}{99}
\bibitem{note3} For 2D Bose gases, time-reversal symmetry can be
spontaneously broken by creating vortex-antivortex pairs below the BKT
transition temperature. However, since the excitation energy increases
logarithmically with the vortex pair size, for temperature not too close to
the transition temperature, vortex pairs are tightly combined and we can
resume a time-reversal symmetry for the coarse-grained wave function.
\end{thebibliography}
\end{document}